\documentclass[conference]{IEEEtran}
\IEEEoverridecommandlockouts
\usepackage{cite}
\usepackage{amsmath}
\usepackage{amsmath,amssymb,amsfonts}
\usepackage{algorithmic}
\usepackage{graphicx}
\usepackage{textcomp}
\usepackage{xcolor}
\def\BibTeX{{\rm B\kern-.05em{\sc i\kern-.025em b}\kern-.08em
    T\kern-.1667em\lower.7ex\hbox{E}\kern-.125emX}}
\begin{document}

\title{Data User-Based Attribute Based Encryption\\
}

\author{\IEEEauthorblockN{Ehsan Meamari~~~~~Hao Guo~~~~~Chien-Chung Shen~~~~~Rui Zhang}
\IEEEauthorblockA{Department of Computer and Information Sciences\\
University of Delaware, U.S.A. \\
\{ehsan, haoguo, cshen, ruizhang\}@udel.edu}}

\maketitle

\begin{abstract}
Attribute Based Encryption (ABE) has emerged as an informtion-centric public key cryptographic system which allows a data owner to share data, according to an access policy, with multiple data users based on the attributes they possess, without knowing their identities. In the original ABE schemes, a central authority administrates the system and issues secret keys to data users based on their attributes and both the owner and users need to trust a specific CA. However, in certain real-world applications, the data users would not trust anyone but themselves. For such situations, we introduce a new decentralization model of ABE, termed Data User-based ABE (DU-ABE), which is managed jointly by the data users. DU-ABE is the first decentralized ABE scheme which replaces the authorities with the data users without employing any other extra entities.
\end{abstract}


\section{Introduction}

For Bob to communicate securely with Alice via public-key encryption, he encrypts a message with Alice's public key (PK), which Alice decrypts with her secret (private) key (SK). Here, Bob (a data owner or DO) knows that it is Alice (a data user or DU) who he would like to communicate with. However, there are situations where a  DO would like to share data securely with (multiple) DUs whose identities are not known. For instance, the President of a university may want to send encrypted data to all the Teacher or Research Assistants in the CIS Department, but does not know the identities of all these eligible students.

In 2005, Sahai and Waters \cite{Sahai-2005-Introducing-ABE} introduced Attribute-Based Encryption (ABE) as the first one-to-many cryptosystem. There are two kinds of ABE: ciphertext policy attribute-based encryption (CP-ABE) and key policy attribute-based encryption (KP-ABE). The first KP-ABE was proposed by Goyal et al. in 2006 \cite{Goyal-2006-First-KP-ABE} and the first CP-ABE by Bethencourt et al. in 2007 \cite{Bethencourt-First-CP-ABE}.

In CP-ABE, a DU receives its SK based on the attributes it possesses. When the DO shares the data securely, the data is encrypted with an access policy. For instance, the University President could share data securely with the policy \{``CIS Department" \& (``Teacher Assistant, '' $||$ ``Research Assistant'')\} so that students with either attribute \{``CIS Department" \& ``Teacher Assistant''\} or \{``CIS Department" \& ``Research Assistant''\}  can decrypt the President's ciphertext. In KP-ABE, DUs' SKs are generated based on an specific access policy. Clearly, CP-ABE is more applicable to practical applications.

The original ABE systems employ a central authority (CA) which issues SKs to DUs. Such a centralized architecture suffers from several issues. Efforts \cite{Hur-First-Paper, Hur-DABE-with-Revocation,Lin-DABE-with-two-Secure-2PC,Kowshalya2014, Zhao-similar-Hur-first-model-but-discussed-2PC-protocol,Li-Decentralizing-both-Models-of-Water-and-Lewko,Chase-first-decentralized-(KP-ABE)-Multiple-authorities-managed-by-CA,Nyamsuren2018,Guo-Multi-Authority-Attribute-Based-Access-Control-with-SC,Sultan2017} have been proposed to decentralize ABE to address issues such as key escrow \cite{Bozovic2012-A-Model-based-on-Chase-first-decentralized-(KP-ABE)-when-DO-defines-list-valid-DUs}, ineligible DUs \cite{Yu:2016:ACA:3090725.3090727}, key exposure \cite{Ramana2015}, forging signatures \cite{Wang2012}, privacy of DUs \cite{Song-Privacy-Preserving-via-blank-token}, and flexibility \cite{Lewko-Decentralizing-Attribute-Based-Encryption}. The main idea of decentralization lies in dividing the responsibilities of the single CA among multiple authorities.

In both ABE and decentralized ABE (DABE), the CA/authorities are assumed to work honestly so that all the DOs and DUs are obligated to trust CA/authorities.
However, since there is no supervision of CA/authorities, there are situations where DUs would not trust CA or authorities. Therefore, the existing DABE models are not capable of handling these situations.

To address the problem of unwillingness to trust CA/authorities, this paper proposes a new DABE scheme termed data user-based ABE (or DU-ABE for short), where the DUs are not obligated to trust CA/authorities. Instead, the DUs themselves take over the responsibilities of CA/authorities. To the best of our knowledge, this is the first scheme where DUs work as the authorities of an ABE system. Although, in several efforts \cite{Song-Privacy-Preserving-via-blank-token, Chase-Multi-Authority-(KP-ABE)-Preserves-Privacy-but-Supports-just-AND-policies-Needs-Too-Much-Communications-among-AAs}, DUs cooperatively work with CA/authorities to facilitate certain features, DUs do not have any role to issue SKs to other DUs. In addition, these models employ both CA/authorities and DUs to operate the system. In contrast, DU-ABE completely removes any other entity except the DUs themselves, so that all the activities are accomplishes by the DUs.

The paper proceeds in Section~\ref{sec:preliminaries} to review the needed background knowledge. In Section \ref{sec:categorizing}, the paper reviews the issues raised by centralized ABE and then categorizes the existing DABE schemes. Section~\ref{sec:DU-ABE} first articulates a new incentive to decentralise ABE and then describes the new decentralised ABE model, DU-ABE. Section~\ref{sec:Conclusion} concludes the paper with future work. 

\section{Preliminaries}
\label{sec:preliminaries}

In this section, we first briefly review some basic cryptographic backgrounds, and then explain the concept of CP-ABE which is adapted to develop our model.


\subsection{Bilinear Maps}

Consider $G_0$ and $G_1$ as two multiplicative cyclic groups, $p$ as the same prime order of both groups, and $g$ as the generator of $G_0$. A map $e: G_0 \times G_0 \rightarrow G_1$, with an efficient algorithm to compute $e(g_1, g_2)$ for all $g_1, g_2 \in G_0$. Then $e$ is called bilinear if it has the following two properties:

1) \textbf{Bilinearity:} For all $g_1, g_2 \in G_0$ and $ a, b \in Z_p$, $e(g_1^a, g_2^b) = e(g_1, g_2)^{ab}$.

2) \textbf{Non-degeneracy:} $e(g, g) \neq 1$.






\subsection{Review of CP-ABE}

As CP-ABE is more applicable to real-world applications than KP-ABE, most of the research has been done to improve CP-ABE. The first applicable CA-ABE scheme was proposed by Bethencourt et al. \cite{Bethencourt-First-CP-ABE} which was then improved by Waters \cite{Waters-CP-ABE} to provide a more efficient scheme. The two models from these two papers are the basic schemes which are used in subsequent papers. Both of these two models consist of the  following four algorithms.

\textbf{Setup}: The CA runs the Setup algorithm by choosing some security parameters and an attribute universe description to generate a PK and a master key (MK). Then, the PK is broadcast and known by every entity in the system, while the MK is known only to the CA. The CA needs the MK to issue SKs to different DUs.

\textbf{KeyGen}: Each DU who wants to decrypt ciphertexts should send a request to the CA to receive the SKs related to DU's attributes. While the CA knows the MK, it can issue the SKs for any legitimate DUs based on DU's attributes by running the KeyGen algorithm.

\textbf{Encrypt}: The DO runs this algorithm to encrypt its data by using the PK. It also needs to determine a specific access policy to decide which DUs should be able to decrypt the ciphertext. Then, the DO shares the ciphertext with the DUs.

\textbf{Decrypt}: The DU receives the ciphertext and decrypts the it by only if he/she possesses enough attributes which are needed based on the mentioned access policy by the DO. The Decrypt algorithm returns an Error when a DU does not possess enough attributes.




\section{Review of Existing DABE Models}
\label{sec:categorizing}

In this section, we first review some of the issues that incentivize the decentralization of ABE (into DABE). We then classify existing DABE models into two categories.

\subsection{Issues with Centralized ABE}

In the original CP-ABE models, there is one CA managing the entire system. Such centralized architectures raise the following issues.

\subsubsection{Key escrow}

In centralized ABE, since CA has the MK of the entire system, a `curious' CA might generate SKs related to all the attributes for itself and decrypt all the ciphertexts \cite{Bozovic2012-A-Model-based-on-Chase-first-decentralized-(KP-ABE)-when-DO-defines-list-valid-DUs}. However, DOs only want to share their data with legitimate DUs, not the `curious' CA. To address this issue, efforts of \cite{Hur-First-Paper, Wang2016-Decentralizing-Waters-model-by-2PC} divided the responsibilities of the CA (of centralized ABE) among multiple authorities so that each authority generates its own MK. To issue an SK, all the authorities should cooperate with each other so that each authority generates one component of the SK from its MK. A DU can form its SK by `combining' all the components of SKs received from authorities. Therefore, as long as the authorities do not collude each other by sharing their MKs with each other, none of the authorities can issue a SK for itself.

\subsubsection{Ineligible DUs}

DOs are obligated to trust the CA and assume that the CA will authenticate the eligible DUs honestly. However, as the CA is not supervised in centralized ABE, the CA might take advantage of its power and issue SKs to ineligible DUs \cite{Yu:2016:ACA:3090725.3090727}. Decentralization mitigates this issue since a (decentralized) authority cannot generate SKs alone as it has to cooperate with other authorities to issue SKs.

\subsubsection{Key exposure}

After running the Setup algorithm and generating the MK, the CA needs to save the MK securely. In practice, an attacker might circumvent the security of CA and steal its MK \cite{Ramana2015}, so that the entire system is compromised. In DABE, the issue of key exposure is mitigated by each authorities having its own MK. An adversary will not be successful until stealing the MK from all the authorities, which is much more complicated.

\subsubsection{Forging signatures}

In centralized ABE, after generating SKs, the CA can save a copy of the SKs to forge the signatures of the DUs who use these SKs as signatures \cite{Wang2012}. In the DABE schemes of \cite{Hur-Decentralized-ABE-for-Military-Networks, Li-Decentralizing-both-Models-of-Water-and-Lewko, Wang2016-Decentralizing-Waters-model-by-2PC}, since the final SK is generated by multiple authorities, none of the authorities has the final SK to forge the signature of a DU.


\subsubsection{Privacy of DUs}

Privacy becomes a concern when the CA uses the attributes of a DU to generate the SK. Song et al. \cite{Song-Privacy-Preserving-via-blank-token} proposed a DABE scheme which consists of the authorities of attribute auditing center (AAC) and key generation center (KGC). DU authenticates itself to AAC to receive a `blind token' which is presented to KGC to generate the SK. In this case, the model preserves privacy as AAC does not know anything about the SK and KGC does not know anything about the attributes of the DU.

\subsubsection{Flexibility}
Centralized ABE is not scalable to scenarios where organizations want to share data according to a policy written over attributes issued across different trust domains. Lewko et al. \cite{Lewko-Decentralizing-Attribute-Based-Encryption} proposed an scalable multi-authority ABE scheme in which a DO can encrypt data for a policy written over attributes issued by different authorities.

\subsection{Classification of the Existing DABE Schemes}

The DABE schemes described above can be broadly classified into the following two categories.

\subsubsection{Authority-level decentralization} One obvious approach to decentralizing ABE is to divide the responsibilities of the CA among authorities, so that each of the authorities has its own MK and authorities cooperatively issue an SK to each DU. Hur et al. \cite{Hur-First-Paper} employed two authorities where they run a secure two-party computation protocol to issue SK to DUs without revealing their MKs to each other. While the scheme in \cite{Hur-First-Paper} is based on the Bethencourt model \cite{Bethencourt-First-CP-ABE}, Li et al. \cite{Li-Decentralizing-both-Models-of-Water-and-Lewko}, based on the Waters model \cite{Waters-CP-ABE}, developed a similar decentralization scheme with two authorities. In addition, Li et al. \cite{Li-Decentralizing-both-Models-of-Water-and-Lewko} proposed a DABE scheme in which a threshold $t$ out of $n$ authorities are needed to collaborate with each other and issue SK to a DU.

In contrast, Lewko et al. \cite{Lewko-Decentralizing-Attribute-Based-Encryption} delegated the responsibility of generating the part of SK for each attribute (or a set of attributes) to one authority, so that authorities do not need to cooperate with one other to generate the final SK.

\subsubsection{DO-level decentralization}

Bozovic et al. \cite{Bozovic2012-A-Model-based-on-Chase-first-decentralized-(KP-ABE)-when-DO-defines-list-valid-DUs} developed a model in which a DO cooperate with multiple authorities to operate a DABE system by explicitly defining the list of recipients who can decrypt the ciphertext. In addition, Sandor et al. \cite{Sandor2019-DO-issue-D-part-other-AAs-issue-SK-for-each-attribute-with-outsourced-decryption} proposed a model in which multiple authorities cooperate with DO during the key generation phase. In this model, it is supposed that DO has a local list of DUs and can authenticate these DUs. Then DO and each of the authorities together issue one part of SK to DU. The final SK is a combination of all the different parts of SK.

Although DO-level decentralization can mitigate some of the issues incurred by centralized ABE, DOs need to know the set of recipients to issue SK parts to them. This assumption is in contrast to the original motivation of ABE in which DOs do not know the DUs in advance.

\section{Data User-based ABE (DU-ABE)}
\label{sec:DU-ABE}

The previous DABE schemes have been developed to mitigate the above mentioned issues associated with centralized ABE schemes. In this section, we describe a new incentive which prompts the development of a new decentralization model of ABE.

\subsection{New Incentive for Decentralization}
\label{sec:practical-applications}

We use the following example to  motivate the new model. In a country where members of the House of Representatives form different committees to makes the nation's laws, each member possesses different attributes, such as member of the ``House of Representatives", member of the ``Agriculture committee," member of the ``Budget committee," and so on. By using ABE, a citizen (i.e., a DO) can share private information with the intended members (i.e., DUs) according to his/her access policy. 

Typically, members of the House of Representatives are either affiliated with different parties or independent. To use a centralized ABE system, all the members of the House of Representative are obligated to agree on a CA. The question is who could be the CA. One simple option is that all the members trust the Speaker to become the CA who operates a centralized ABE system to issues SKs to members. However, the issue is that the Speaker is probably from the majority party who will not be trusted by either the members of the other parties or independent. In such a scenario, issues of key escrow, ineligible DUs, key exposure, and forging signature described above will arise.

The other option is to have a subset of members as authorities to run a DABE system, where all the members need to trust the authorities and assume that the authorities will not collude with each other. However, there are independent members who will not trust any authorities.

To address this issue, we introduce DU-ABE which is designed to handle such a situation. In DU-ABE, all the DUs possess attributes, and run a DU-ABE system cooperatively to issue SKs to themselves. Members of a specific committee, each having his/her own MK, issue SKs to one another. Therefore, in DU-ABE, all the DUs are authorities themselves where they do not need to trust one another.

\subsection{Model of DU-ABE}

In DU-ABE, all the DUs together have the full control of the cryptosystem, where DUs generate and save their respective MKs. Cooperatively, they issue SKs to themselves without leaking their respective MKs to one another.  DU-ABE consists of the following five algorithms:

\textbf{Global Setup} $(\lambda)\rightarrow$ GP

All the DUs need to communicate with one another to agree upon the security parameters ($\lambda$) as the input to the Global Setup algorithm which outputs the global parameters (GP) for the system.

\textbf{Authority Setup} (GP) $\rightarrow$ MK, PK

Every group of DUs who possess the same attribute run the Authority Setup algorithm together, which inputs the GP of the system and outputs both an MK and a PK for that attribute for each DU. The DUs of different groups broadcast the PKs but save the MKs securely. So, the system has a distinct MK and PK per attribute for each DU.

\textbf{Encryption} ($M, (A, \rho)$, GP, \{PK\}) $\rightarrow$ CT

A DO runs the encryption algorithm which inputs a message $M$, an access matrix $(A, \rho)$, which is generated based on the access policy, the GP of the system, and also the set of PKs of the relevant set of attributes specified in access matrix. The encryption algorithm outputs a ciphertext (CT), which only a DU who possesses the specified attributes, can decrypt the CT and retrieve the message $M$.

\textbf{KeyGen} (GID, GP, MK) $\rightarrow$ SK

When a specific DU wants to obtain an SK related to one attribute, it needs to communicate with all the other DUs who have the same attribute. This DU is one of the DUs who are the authorities of this attribute. Then, these DUs run the key generation algorithm which takes GID as an identity for the DU, $i$ related to the attribute, general parameter (GP) and MK of the DUs, and outputs the SK to the legitimate DU.

\textbf{Decrypt} (CT, GP, SK) $\rightarrow M$

A DU needs to run the decryption algorithm to decrypt the ciphertext CT. Th decryption algorithm takes CT, global parameter (GP), and secret key SK related to a DU with its specific GID. If the DU possesses enough attributes, it can decrypt the CT to access the message $M$.

\subsection{Construction of DU-ABE}

We develop DU-ABE based on the Lewko model \cite{Lewko-Decentralizing-Attribute-Based-Encryption} based on the following algorithm.

\textbf{Global Setup} ($\lambda) \rightarrow$ GP

In the global setup algorithm, all the DUs agree on a bilinear group $G$ of order $N = p_{1}p_{2}p_{3}$, a production of three prime numbers, and a generator $g_{1}$ of $G_{p1}$. In addition, DUs choose a hash function $H: \{0, 1\}^* \longrightarrow G$ that maps GID of DUs to elements of $G$. Then, GP = $\{g_1, H\}$ is published.

\textbf{Authority Setup} (GP) $\rightarrow$ PK, MK

We suppose that each attribute $i$ is managed by a group of $n_i$ DUs who possess attribute $i$. Each DU$_{i,j}$, as the $j^{th}$ data user who possesses attribute $i$, should choose two random exponents $\alpha_{i,j}, \beta_{i,j} \in Z_N$ and keep MK$_{i,j} = \{\alpha_{i,j}, \beta_{i,j}\}$ as its master key. Then DU$_{i,j}$ publishes its public key PK$_{i,j} = \{e(g_1, g_1)^{\alpha_{i,j}}, g_1^{\beta_{i,j}}\}$. The PK of attribute $i$ (PK$_i$) can be generated by multiplying the PKs of all $n_i$ DUs who manage attribute $i$ as follows:

PK$_i$ = $\prod_{k=1}^{n_i}$ PK$_{i,k} = \{e(g_1, g_1)^{\sum_{k=1}^{n_i} \alpha_{i,k}}, g_1^{\sum_{k=1}^{n_i} \beta_{i,k}}\}$

\textbf{KeyGen} (GID, GP, \{MK$_{i,j}$\}) $\rightarrow$ SK$_{i,\mathrm{GID}}$

When a DU with a specific GID wants to receive its secret key SK$_{i,\mathrm{GID}}$ for attribute $i$, it should send a request to the other $n_i-1$ DUs who possess attribute $i$. Each of $n_i$ DUs issue $ {g_1}^{\alpha_{i,j}} H(\mathrm{GID})^{\beta_{i,j}}$ to the requesting DU and the final SK$_{i,\mathrm{GID}}$ is as follows.

SK$_{i,\mathrm{GID}}$=$\prod_{k=1}^{n_i} {g_1}^{\alpha_{i,k}} H(\mathrm{GID})^{\beta_{i,k}}$

As the public parameters, the public keys, and the final generated are the same as those of the Lewko model \cite{Lewko-Decentralizing-Attribute-Based-Encryption}, both encryption and decryption algorithms remain the same.





\begin{figure}[h]
\centering
\includegraphics[width=0.35\textwidth]{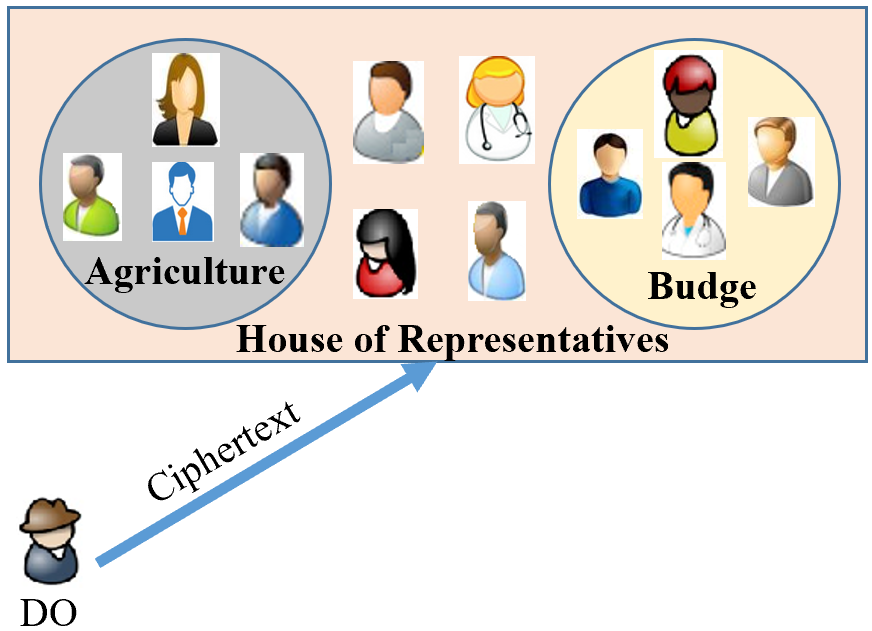}
\caption{An DU-ABE Example}
\label{fig:DU-ABE}
\end{figure}

For instance,  Fig.~\ref{fig:DU-ABE} depicts a simplified example of a House of Representatives, where all the members possess attribute ``House of Representatives,'' and some of them possess attributes ``agriculture'' or ``Budget.'' First, all the members need to agree on security parameters $\lambda$ and run the Global Setup algorithm. Then all the members run the Authority Setup algorithm for attribute ``House of Representatives,'' and each member chooses an MK and publishes a PK for this attribute. Afterward, all the members issue a part of SK to one another for this attribute.

For attribute ``Budget,'' only four members run the Authority Setup algorithm and choose MK and publish PK. Then these four members issue parts of SK to one another. The same activities for attribute ``Agriculture.'' Using encryption algorithm, a DU such as the Secretary of the Treasury, can encrpt data with access policy {``House of Representatives'' \& ``Budget''} so that only the member of the House of Representatives who are also members in the Budget committee can decrypt the ciphertext.



\section{Analyses of DU-ABE}

\subsection{Security}

DUs might collude with each other by combining their SKs to decrypt a ciphertext, which otherwise cannot be decrypted by using individual SK. DU-ABE is secure against this kind of collusion attack, since the issued SKs are related to specific GIDs for different DUs. Therefore, a combined SK with different GIDs could not be used for decryption.

In DU-ABE, as long as at least one DU works trustworthy, the system would be secure. Therefore, each DU need to think about himself and work honestly. If a DU does not collude with other DUs, he/she could be sure that other DUs cannot issue a SK and launch key escrow nor ineligible DU attacks.

\subsection{Limitations}

In contrast to all the other ABE and DABE systems, in DU-ABE, the DUs need to know each other before bootstrapping the system, as DUs collaboratively accomplish all the functions.

In addition, all the DUs need to communicate with one another to issue SKs to each DU. Although this incurs high communication overhead, this key generation phase occurs only once when the system boots up.

\section{Conclusion}
\label{sec:Conclusion}

In this paper, we introduced the concept of Data Users-based ABE (DU-ABE) as a new decentralized ABE model. As discussed, in situations where the DUs cannot trust each other or any other authorities, they bootstrap a DU-ABE system to manage the entire system by themselves. In the future work, we plan to add features, such as revocation to accommodate dynamic membership, and formally prove the security of DU-ABE.

\bibliographystyle{IEEEtran}
\bibliography{Reference}

\end{document}